\documentclass[traditabstract]{aa} 
\usepackage{graphicx}
\usepackage{txfonts}
\usepackage{natbib}
\bibpunct{(}{)}{;}{a}{}{,} 
\begin{document}
   \title{The VIMOS Ultra Deep Survey First Data Release: spectra and spectroscopic redshifts of 698 objects up to $z_{spec}$$\sim$$6$ in CANDELS
	  \thanks{Based on data obtained with the European 
	  Southern Observatory Very Large Telescope, Paranal, Chile, under Large
	  Program 185.A--0791. }
}

\author{L. A. M.~Tasca\inst{1}
\and O.~Le F\`evre\inst{1}
\and B.~Ribeiro\inst{1}
\and R.~Thomas\inst{1}
\and C. Moreau \inst{1}
\and P.~Cassata\inst{5}
\and B.~Garilli\inst{6}
\and V.~Le Brun\inst{1}
\and B.~C.~Lemaux \inst{1}
\and D.~Maccagni\inst{6}
\and L.~Pentericci\inst{7}
\and D. Schaerer\inst{2,3}
\and E.~Vanzella\inst{4}
\and G.~Zamorani \inst{4}
\and E.~Zucca\inst{4}
\and R.~Amorin\inst{7}
\and S.~Bardelli\inst{4}
\and L.~P.~Cassar\`a\inst{6}
\and M.~Castellano\inst{7}
\and A.~Cimatti\inst{8}
\and O.~Cucciati\inst{8,4}
\and A.~Durkalec\inst{1,21}
\and A.~Fontana\inst{7}
\and M.~Giavalisco\inst{9}
\and A.~Grazian\inst{7}
\and N.P.~Hathi\inst{1}
\and O.~Ilbert\inst{1}
\and S.~Paltani\inst{10}
\and J.~Pforr\inst{1}
\and M.~Scodeggio\inst{6}
\and V.~Sommariva\inst{8,7}
\and M.~Talia\inst{8}
\and L.~Tresse\inst{1}
\and D.~Vergani\inst{11,4}
\and P.~Capak\inst{12}
\and S.~Charlot\inst{13}
\and T.~Contini\inst{14}
\and S.~de la Torre\inst{1}
\and J.~Dunlop\inst{15}
\and S.~Fotopoulou\inst{10}
\and L.~Guaita\inst{7}
\and A.~Koekemoer\inst{16}
\and C.~L\'opez-Sanjuan\inst{17}
\and Y.~Mellier\inst{13}
\and M.~Salvato\inst{18}
\and N.~Scoville\inst{19}
\and Y.~Taniguchi\inst{20}
\and P.W. Wang\inst{1}
}

\institute{Aix Marseille Universit\'e, CNRS, LAM (Laboratoire d'Astrophysique de Marseille) UMR 7326, 13388, Marseille, France
\and
Geneva Observatory, University of Geneva, ch. des Maillettes 51, CH-1290 Versoix, Switzerland
\and
Institut de Recherche en Astrophysique et Plan\'etologie - IRAP, CNRS, Universit\'e de Toulouse, UPS-OMP, 14, avenue E. Belin, F31400
Toulouse, France
\and
INAF--Osservatorio Astronomico di Bologna, via Ranzani, 1 - 40127, Bologna
\and
Instituto de Fisica y Astronom\'ia, Facultad de Ciencias, Universidad de Valpara\'iso, Gran Breta$\rm{\tilde{n}}$a 1111, Playa Ancha, Valpara\'iso Chile
\and
INAF--IASF Milano, via Bassini 15, I--20133, Milano, Italy
\and
INAF--Osservatorio Astronomico di Roma, via di Frascati 33, I-00040, Monte Porzio Catone, Italy
\and
University of Bologna, Department of Physics and Astronomy (DIFA), V.le Berti Pichat, 6/2 - 40127, Bologna
\and
Astronomy Department, University of Massachusetts, Amherst, MA 01003, USA
\and
Department of Astronomy, University of Geneva
ch. d'\'Ecogia 16, CH-1290 Versoix, Switzerland
\and
INAF--IASF Bologna, via Gobetti 101, I--40129,  Bologna, Italy
\and
Department of Astronomy, California Institute of Technology, 1200 E. California Blvd., MC 249-17, Pasadena, CA 91125, USA
\and
Institut d'Astrophysique de Paris, UMR7095 CNRS,
Universit\'e Pierre et Marie Curie, 98 bis Boulevard Arago, 75014
Paris, France
\and
Institut de Recherche en Astrophysique et Plan\'etologie - IRAP, CNRS, Universit\'e de Toulouse, UPS-OMP, 14, avenue E. Belin, F31400
Toulouse, France
\and
SUPA, Institute for Astronomy, University of Edinburgh, Royal Observatory, Edinburgh, EH9 3HJ, United Kingdom
\and
Space Telescope Science Institute, 3700 San Martin Drive, Baltimore, MD 21218, USA
\and
Centro de Estudios de F\'isica del Cosmos de Arag\'on, Teruel, Spain
\and
Max-Planck-Institut f\"ur Extraterrestrische Physik, Postfach 1312, D-85741, Garching bei M\"unchen, Germany
\and
Department of Astronomy, California Institute of Technology, 1200 E. California Blvd., MC 249--17, Pasadena, CA 91125, USA
\and
Research Center for Space and Cosmic Evolution, Ehime University, Bunkyo-cho 2-5, Matsuyama 790-8577, Japan 
\and
National Centre for Nuclear Research, Astrophysics Division, Hoza 69, 00-681, Warszawa, Poland \\ \\
             \email{lidia.tasca@lam.fr}
}

   \date{Received ...; accepted ...} 

%
 
  \abstract{
This paper describes the first data release (DR1) of the VIMOS Ultra Deep Survey (VUDS). The DR1 includes all low--resolution spectroscopic data obtained in 276.9 arcmin$^2$ of the CANDELS--COSMOS and CANDELS--ECFDS survey areas, including accurate spectroscopic redshifts $z_{spec}$ and individual spectra obtained with VIMOS on the ESO--VLT. A total of 698 objects have a measured redshift, with 677 galaxies, two type--I AGN and a small number of 19 contaminating stars. The targets of the spectroscopic survey are selected primarily on the basis of their photometric redshifts to ensure a broad population coverage. About 500 galaxies have $z_{spec}>2$, 48 with $z_{spec}>4$, and the highest reliable redshifts reach beyond $z_{spec}=6$. This dataset approximately doubles the number of galaxies with spectroscopic redshifts at $z>3$ in these fields. We discuss the general properties of the sample in terms of the spectroscopic redshift distribution, the distribution of Lyman$-\alpha$ equivalent widths, and physical properties including stellar masses M$_{\star}$ and star formation rates (SFR) derived from spectral energy distribution fitting with the knowledge of the knowledge of $z_{spec}$. We highlight the properties of the most massive star-forming galaxies, noting the large range in spectral properties, with Lyman--$\alpha$ in emission or in absorption, and in imaging properties with compact, multi-component or pair morphologies. We present the catalogue database and data products. All data are publicly available and can be retrieved from a dedicated query--based database available at \texttt{http://cesam.lam.fr/vuds}. 
}


\authorrunning{Lidia A. M. Tasca et al.}
\titlerunning{VIMOS Ultra Deep Survey DR1}

   \maketitle


\section{Introduction}

Deep uniformly targeted galaxy surveys are reliably uncovering the average properties of distant galaxy populations, a primary source of information to consolidate a galaxy formation and evolution scenario. Spectroscopic surveys play a key role as they provide accurate redshifts together with a wealth of spectral properties including continuum, emission or absorption line features and indexes, which allow to fully characterize the physical properties of galaxies. An important goal of these surveys is to provide robust properties of galaxy populations and their distributions around mean values from volume--complete samples, and also enable serendipitous discoveries and studies of rare populations.

A number of spectroscopic galaxy surveys were performed in the past years, reaching increasingly higher redshifts and larger volumes. The local universe is well surveyed by the staged Sloan Digital Sky Survey (SDSS) with more than a million galaxies observed at redshifts $z\sim0.1-0.7$, as presented in the latest DR12 data release \citep{Alam:2015,Reid:2015}. At redshifts $\sim1$ several spectroscopic surveys have sampled large volumes of the universe with  $\sim$10\,000 to $\sim$100\,000 galaxies, VVDS--Wide \citep{LeFevre:05,LeFevre:13a},  DEEP2 \citep{Davis:2003}, zCOSMOS \citep{Lilly:07}, VIPERS \citep{Guzzo:14}, providing a reliable description of the galaxy properties in relation to their local environment. At redshifts $z>2$ spectroscopic surveys have been more limited in scope with several thousand galaxies identified at $z\sim2-4$ \citep{Steidel:03,Bielby:2011,LeFevre:13a,Kriek:2015,Momcheva2015}, and only a few hundred at $z>4$ \citep{Vanzella:2009,Stark:2009,LeFevre:13a,Pentericci2014}. At these high redshifts the main reference samples discussed in the literature are mostly based on photometric redshift estimates obtained from SED fitting of a set of broad and medium band filters \citep[e.g.][]{Ilbert:06,Ilbert:09,Ilbert:10,Ilbert:13}, are selected using the Lyman-break technique \citep[e.g. ][]{Bouwens:14}, or result from the identification of excess flux in narrow band filters selecting high--z Lyman--$\alpha$ emitters \citep[e.g. ][]{Rhoads:2003,Ouchi:2008}.

Data releases from spectroscopic surveys are a fundamental step to provide access to a documented data set for a wide community. They enable a wide range of science investigations and provide a basis to establish a galaxy evolution scenario that needs to be accurately reproduced by theoretical predictions and numerical simulations. The availability of a documented reference for such data releases greatly aids this process.
The galaxy samples accessed from data releases provide an observational reference upon which our view of the universe is built. SDSS has produced 12 data releases \citep[DR12; ][]{Alam:2015,Reid:2015}, and data release papers can be found for most major spectroscopic surveys including VVDS \citep{LeFevre:13a}, DEEP2 \citep{DEEP2:2013}, zCOSMOS--bright \citep{Lilly:2009}, and VIPERS \citep{Garilli:2014}. The interactions of users with large datasets are easily facilitated by the availability of databases, where query-based interfaces allow advanced data mining.

In this paper we present the first data release (DR1) of the VIMOS Ultra Deep Survey (VUDS). VUDS is conceived to alleviate some of the limitations related to photometric samples by assembling a large sample of galaxies {\it with spectroscopic redshifts} up to some of the highest redshifts that can be reached with an 8m--class telescope \citep{LeFevre:15}. The primary goal of VUDS is to provide an accurate view of star--forming galaxies in the distant universe at redshifts $z>2$ and up to $z\sim6$, as seen from a UV rest-frame perspective. With a large sample assembled over a large volume and star--forming galaxies dominating galaxy counts it becomes possible to build a robust statistical description of the galaxy population at a key time in galaxy evolution, leading to an improved understanding of galaxy assembly at early cosmic times. Targets in VUDS are selected based on their photometric redshifts supplemented with colour selection and then observed with long integrations with the VIMOS multi--slit spectrograph on the VLT. With $\sim$10\,000 objects targeted in one square degree built from three separate fields to minimize cosmic variance, VUDS is today the largest spectroscopic survey of the universe at these high redshifts \citep{LeFevre:15}. 

The availability to a broad community of spectra of high redshift galaxies in well studied extragalactic fields is essential for an improved knowledge of galaxy evolution. 
The VUDS--DR1 presented in this paper consists of 698 spectroscopic redshift measurements and calibrated one--dimensional spectra in 276.9 arcmin$^2$ of the COSMOS and ECDFS areas of the CANDELS survey \citep{Grogin:2011,Koekemoer:11}, as well as associated physical quantities including stellar masses and star--formation rates. 
The VUDS data in the COSMOS and ECDFS regions of the HST--CANDELS survey offers the opportunity to couple the exquisite HST imaging data with the largest deep spectroscopy sample available to date.
 
We present the VUDS survey in CANDELS in Section \ref{survey}, and the observations in Sect.\ref{obs}. The general properties of the sample are discussed in Sect.\ref{properties}, and some remarkable galaxies in this sample are presented in Sect.\ref{objects}. The content of the VUDS data release number 1 is described in Sect.\ref{release}. A summary is provided in Sect.\ref{summary}.

When quoting absolute quantities we use a cosmology with $H_0=70~km~s^{-1}~Mpc^{-1}$, $\Omega_{0,\Lambda}=0.7$ and $\Omega_{0,m}=0.3$. 
All magnitudes are given in the AB system.


\section{The VUDS survey in the CANDELS fields}
\label{survey}

VUDS is a spectroscopic survey targeting  $\approx$10\,000 galaxies to study galaxy evolution in the redshift range $2<z<6+$.  We provide a summary of the survey below and the reader is referred to  \citet{LeFevre:15} for a detailed description.

Prior to spectroscopy, targets are primarily selected from their photometric redshifts. Photometric redshifts $z_{phot}$ are computed using available broad--band visible and infrared photometry, as well as medium--band photometry in the COSMOS field, and fitting the spectral energy distribution (SED) using the Le Phare software \citep{Ilbert:06}. We retain as primary targets all objects at $i_{AB}\leq25$ which satisfy $z_{phot}+1\sigma \geq 2.4$, representing more than 90\% of the sample. Objects which verify Lyman--break galaxy colour selection criteria in the ugr, gri and riz colour--colour diagrams \citep[see criteria in ][]{Bouwens:07}, and which are not already selected by their photometric redshift, are added to the target list; this concerns about 10\% of the final targets.

The VIMOS field of view is 224 arcmin$^2$ covered by four spectrograph channels with external dimensions $14 \times 16$ arcmin$^2$ and a typical footprint as identified in \cite{LeFevre:05}, with the four channels separated by sky areas without VIMOS field coverage. We define the area of this data release as the area in common between the area covered by the CANDELS H--band observations with the WFC3 camera on--board the Hubble Space Telescope \citep{Koekemoer:2011}, and the different VUDS pointings. 

The VUDS--CANDELS data release concerns two of the most observed extragalactic fields: COSMOS at $\alpha_{2000}=10$h$00$m$28.2$s, $\delta_{2000}=02\deg21\arcmin36\arcsec$ and ECDFS at $\alpha_{2000}=03$h$32$m$30$s, $\delta_{2000}=-27\deg48\arcmin00\arcsec$. 

The COSMOS field is rooted on the original HST--COSMOS survey covering about 2 square degrees \citep{Scoville:07b,Koekemoer:07} with a wealth of multi--wavelength optical and infrared photometry \citep[e.g.][]{Taniguchi:07,Ilbert:09,McCracken:12}. The COSMOS field has been the focus of extensive redshift survey campaigns with the VLT \citep[e.g. zCOSMOS survey,][\& Lilly et al. in prep.]{Lilly:07,Lilly:09}, Keck (P. Capak, P.I.) and Subaru \citep{Silverman2015}. In the 163.4 arcmin$^2$ CANDELS area of the COSMOS field, about 1000 galaxies have spectroscopic redshifts from the literature prior to VUDS, mainly at low redshifts but including about 300 galaxies with $2<z<3$.

The ECDFS covers $30\times30$ arcmin$^2$ and has been the target of a number of spectroscopic observations at the VLT \citep[e.g.][]{OLF04,Vanzella08,Balestra10} as well as Magellan \citep{Cooper12}, totalling over 7000 spectroscopic redshifts at various depths and in various redshift ranges. In the CANDELS 113.5 arcmin$^2$ area of the ECDFS about 2400 spectroscopic redshifts are available from the literature prior to VUDS, including about 500 with $z>2$.  

The VUDS--CANDELS release presented here significantly increases the number of available spectroscopic redshifts at $z>2$ in two of the most observed extragalactic fields as described below.


\section{Observations}
\label{obs}

Multi--slit spectroscopy is conducted with the VIsible Multi--Object Spectrograph (VIMOS) on the ESO--VLT unit number 3 "Melipal" \citep{LeFevre:03}. Each VIMOS pointing is observed for about 14 hours in the LRBLUE grism with a wavelength coverage from 3600 to 6700\AA, and another 14 hours with the LRRED grism covering 5500 to 9350\AA. Each of the LRBLUE and LRRED observations cumulate about 40 individual exposures of $\sim20$min integrations each. To remove most of the small scale detector pattern, the telescope is offset between each exposure to move objects along the slits with a repeated offset pattern -1.0, -0.5, 0, +0.5, +1.0 arcsecond from the nominal pointing position.

\begin{figure*}
   \centering
   \includegraphics[width=15cm]{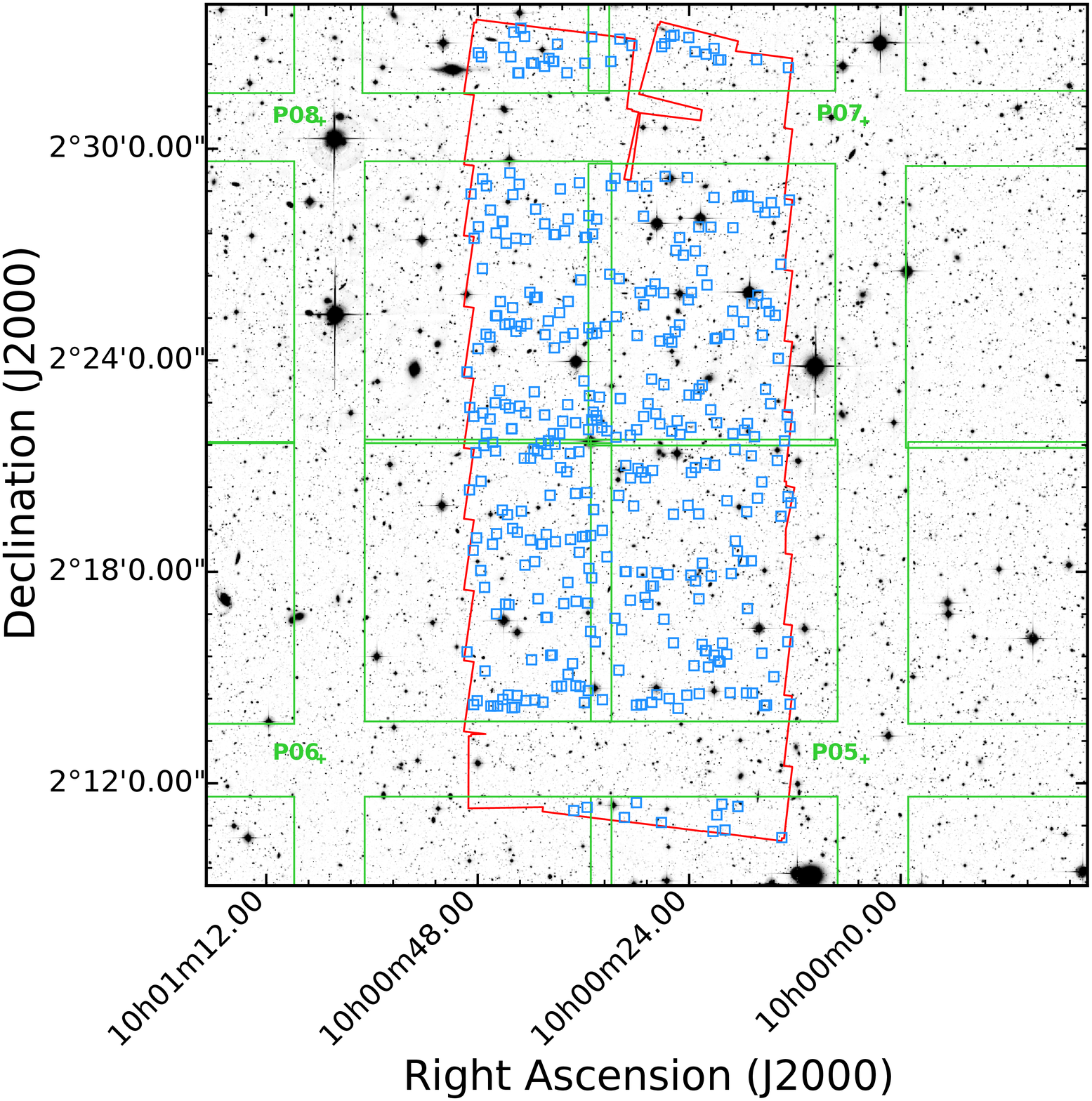}
      \caption{The VUDS coverage of the CANDELS H--band F160W area in the COSMOS field. The CANDELS area is indicated by the red polygon, and the 4--quadrant footprint of the 4 overlapping VIMOS pointings is overlaid as the green rectangles with the pointing number identified. Because of the geometry of the VUDS footprint on the sky, regions corresponding to the $\sim$2 arcmin gap between VIMOS channels did not receive spectroscopic observations as it is apparent in the lower and upper parts of the CANDELS area. The 384 galaxies with VUDS redshifts falling in the area in common between CANDELS and VUDS are plotted as the blue squares.}  
         \label{COSMOS_field}
   \end{figure*}

\begin{figure*}
   \centering
   \includegraphics[width=15cm]{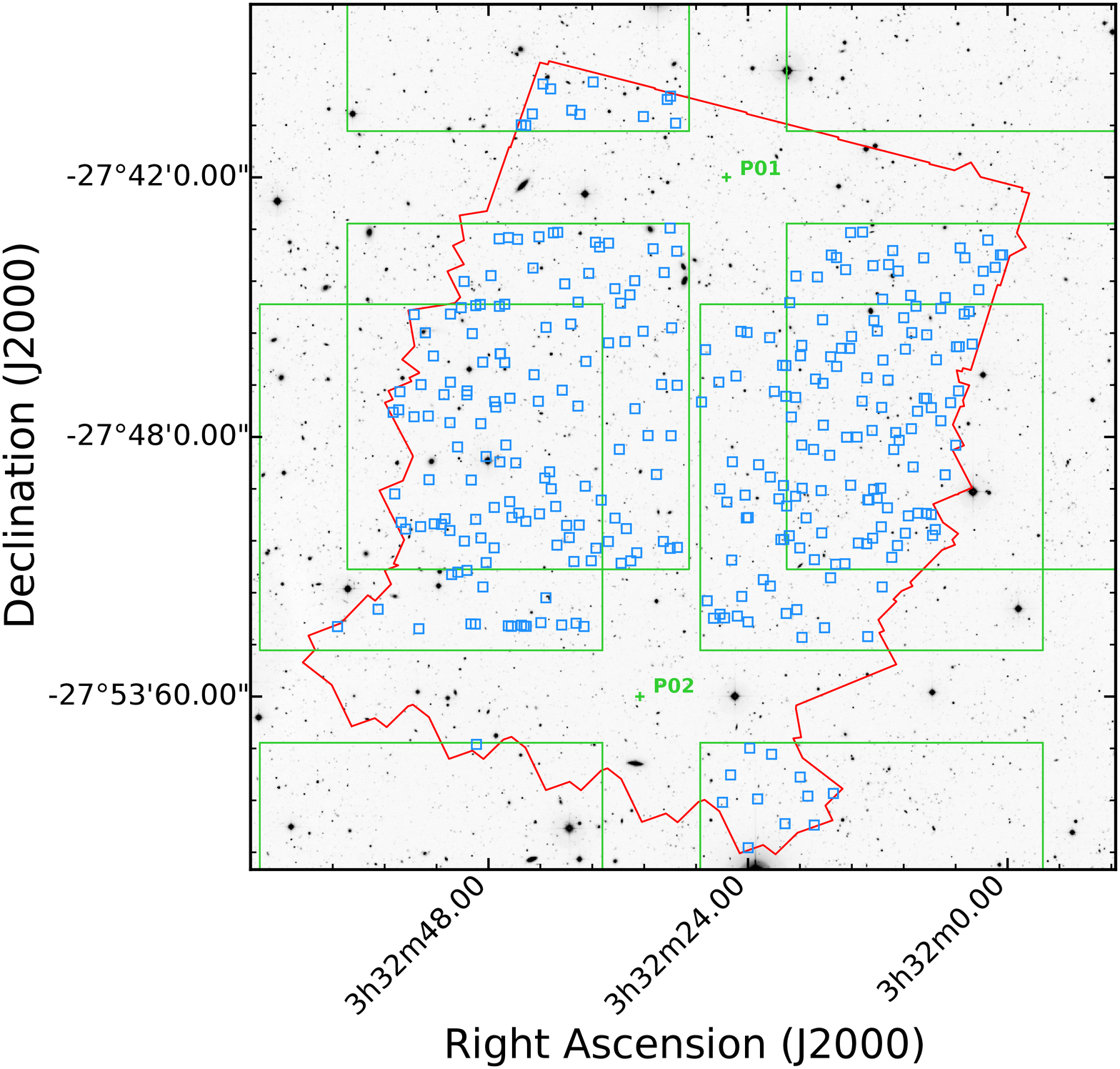}
      \caption{The VUDS coverage of the CANDELS H--band F160W area in the ECDFS field. The CANDELS area is indicated by the red polygon, and the 4--quadrant footprint of the 2 overlapping VIMOS pointings is overlaid as the green rectangles with the pointing number identified. Because of the geometry of the VUDS footprint on the sky, regions corresponding to the $\sim$2 arcmin gap between VIMOS channels did not receive spectroscopic observations as it is apparent in the lower and upper parts of the CANDELS area. The 314 galaxies with VUDS redshifts falling in the area in common between CANDELS and VUDS are plotted as the blue squares.}  
         \label{ECDFS_field}
   \end{figure*}

The VIMOS pointings relevant for this release and associated observing conditions are listed in Table \ref{pointings_candels}. In the COSMOS field eight VIMOS quadrants in four separate VIMOS pointings fall in the CANDELS area (Figure \ref{COSMOS_field} and Table \ref{pointings_candels}). The total area in common between VUDS--COSMOS and CANDELS--COSMOS is 163.43 arcmin$^2$, excluding the regions where VIMOS did not observe \citep[i.e. in the sky area which is blocked from view between the four channels of the VIMOS focal plane][]{LeFevre:05}. The CANDELS area in ECDFS was observed with seven quadrants coming from two VIMOS pointings and include 113.49 arcmin$^2$ of the CANDELS--ECDFS area (Figure \ref{ECDFS_field} and Table \ref{pointings_candels}). 
All observations were executed in the ESO--VLT service mode, requesting excellent observing conditions with seeing less than 1 arcsec FWHM, photometric quality, and dark time with moon distance larger than 60 degrees if present. The process of quality control associated to the service mode observations resulted in an excellent and homogeneous set of observations with seeing in the range FWHM=0.5-1 arcsec. 

Data reduction follows well established protocols for multi--slit observations \citep[e.g.][]{LeFevre95,LeFevre:05,LeFevre:15}. Each observed slit of the multi--slit masks and each of the individual $\sim40$ exposures is processed in the following way using the VIPGI environment \citep{Scodeggio:05}. First, 2D spectrograms are extracted, wavelength calibrated using He+Ar spectra, and sky subtraction is performed at each wavelength using a low--order polynomial fit along the slit profile. The spectra are registered to the same relative position in the slit to take into account the offsets applied during the observations, and then stacked using a
sigma--clipped average which removes most of the instrumental signatures. Spectra are then placed in the FK5 celestial coordinates grid. 
The spectrum associated to the primary target on which the slit has been positioned is identified, and other spectra traces or single emission lines objects (e.g. faint Ly$\alpha$ emitters) are associated to photometric catalogue entries (secondary target, spectroscopic flag 2X, see Sect. \ref{sect_data}), or categorized as spectroscopic emitters without photometric counterpart (flag -2X).
 
One--dimensional spectra are then extracted using optimal weighting based on a Gaussian fit of the 1D profile of each object projected along the dispersion axis, and are sky--corrected as well as flux calibrated using calibrations performed at regular intervals on spectrophotometric standards. The flux calibration assumes that each observing sequence (observing block) is obtained under photometric conditions, which is generally the case for observations in service mode forced to comply to strict observations requirements. This sequence is applied to both the LRBLUE and LRRED grating observations. The LRBLUE and LRRED spectra are then joined to build the final 1D spectrum for each object covering the wavelength range 3600 to 9350\AA \space after scaling based on the average flux in the overlap wavelength range $5500 < \lambda < 6700$\AA \space in common between the LRBLUE and LRRED. Each 1D spectrum is associated with the nearest object in ($\alpha_{2000},\delta_{2000}$) coordinates in the parent target catalogue. A further joint visual examination of photometric images and 2D--spectra is performed to sort out any ambiguity in associating a photometric image to a trace in a 2D--spectrum. 

All 1D and 2D spectra are visually examined to assign a redshift, following a method extensively used for high redshift galaxy surveys as described in \cite{LeFevre:05}, \cite{Lilly:07}, \cite{LeFevre:13a}, and \cite{LeFevre:15}. The method makes use of the EZ redshift measuring package \citep{Garilli:10} which cross--correlates observed spectra with empirical galaxy, AGN, and stellar templates. 

At the end of this process each object observed has flux and wavelength calibrated 2D and 1D spectra associated to it and gets assigned a spectroscopic redshift z$_{spec}$ and a reliability flag z$_{flag}$ (see Section \ref{sect_data} and Table \ref{flags_list} for a description of the spectroscopic flag system).

\begin{table*}
\centering                       
\caption{List of VUDS pointings covering the CANDELS--ECDFS and CANDELS--COSMOS}
\label{pointings_candels}    
\begin{tabular}{ccc}       
\hline\hline                
            Field      &  $\alpha_{2000}$ & $\delta_{2000}$            \\
                       &                  &                  \\
\hline                       
            COSMOS--P05 &  10h00m04.08s & $+02\deg12\arcmin41.4\arcsec$    \\
            COSMOS--P06 &  10h01m05.76s & $+02\deg12\arcmin41.4\arcsec$    \\
            COSMOS--P07 &  10h00m04.08s & $+02\deg30\arcmin46.7\arcsec$    \\
            COSMOS--P08 &  10h01m05.76s & $+02\deg30\arcmin46.7\arcsec$   \\ 
\hline
	    ECDFS--P01  &  03h32m25.99s & $-27\deg41\arcmin59.9\arcsec$   \\
	    ECDFS--P02  &  03h32m34.00s & $-27\deg53\arcmin59.9\arcsec$   \\
\hline                                 
\end{tabular}
\end{table*}

\section{General properties of galaxies in VUDS--CANDELS}
\label{properties}

\subsection{Redshift distribution}
\label{nz_dist}

\begin{figure}
   \centering
   \includegraphics[width=\hsize]{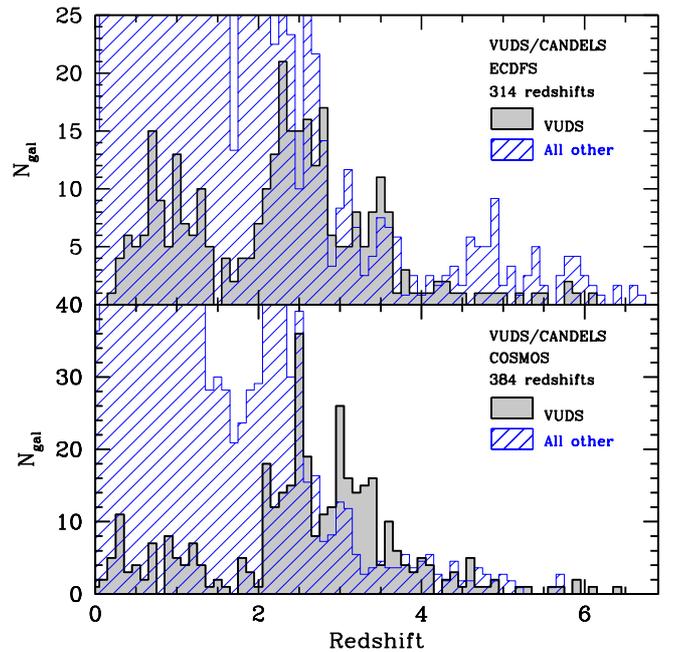}
      \caption{Redshift distribution in the VUDS--CANDELS COSMOS (bottom) and in the VUDS--CANDELS ECDFS (top) fields. All galaxies with a spectroscopic redshift measurement available in these two fields are presented. The light grey histograms are for VUDS sources while the blue diagonal shaded histograms are for measurements obtained by other surveys.}  
         \label{histz}
   \end{figure}

The redshift distribution in each of the VUDS--CANDELS COSMOS and VUDS--CANDELS ECDFS fields is shown in Figure \ref{histz}. The redshift distributions show a number of significant peaks as the lines of sight in each of the fields crosses over the large--scale structure distribution. 

Several peaks in the N(z) distribution are associated with documented structures in the literature and support the existence of these over--densities. In the COSMOS field the two most prominent peaks in the VUDS N(z) are at $z\sim2.45$ and $z\sim2.9$. The strong over--density at $z\sim2.45$ was recently pointed out by \cite{Chiang15} and identified by \cite{Casey15} as a proto--cluster including seven star--bursting submillimeter--luminous galaxies and five AGN. In the COSMOS--CANDELS area alone we find 22 galaxies with $2.45 \leq z_{spec} \leq 2.5$. This appears to be a massive and complex structure comprised of two or three main components which will be further investigated in \citet{Lemaux:16}. The peak at $z\sim2.9$ corresponds to a surrounding structure associated to the proto--cluster identified from the full VUDS data in COSMOS by \cite{Cucciati14} who claim a possible signature of the intra--cluster medium (Ly$\alpha$ absorption at the redshift of the proto--cluster in the spectra of background galaxies).  In the ECDFS two peaks are identified beyond $z=2$ at $z\sim2.9$ and $z\sim3.5$, this latter peak possibly connected to the overdensity reported in \cite{Kang09}. From the redshift distributions in the CANDELS areas and the full VUDS survey area it appears that VUDS does sample a large number of over--dense regions as a result of the large volume sampled and target selection strategy.  

A complete search and 3D characterization of large--scale structures in the full VUDS sample will be presented elsewhere.

\subsection{Synthetic model fitting}
\label{SED}

To derive associated physical parameters for all spectroscopic objects encompassed in VUDS--DR1, e.g., stellar masses, mean luminosity--weighted
stellar ages, SFRs, and rest-frame absolute magnitudes, we utilized the package Le Phare\footnote{http://cfht.hawaii.edu/$\sim$arnouts/LEPHARE/lephare.html}
\citep{Arnouts:99,Ilbert:06,Ilbert:09}. 

The $FUV, NUV, u^{\star}, B, V, g+, r+, i', i+, z+, J, K_s, [3.6], [4.5]$ photometry in the
COSMOS field was drawn from a variety of surveys covering the field \citep{Capak:07,Sanders:2007,Taniguchi:07,Zamojski:07,McCracken:12}.
The process of PSF--homogenizing all UV/optical/ground--based NIR images, source detection, the inclusion of the \emph{Spitzer}/IRAC data,
and the conversion of all magnitudes to 'pseudo--total' magnitudes is described in detail in \citet{Ilbert:13}. The [5.8] and [8.0] 
cryogenic \emph{Spitzer}/IRAC channels were not included in this fitting because offsets of 1--2.5 mags were noticed relative to the best-fit templates
excluding these bands for $\sim$40\% of the objects in VUDS--DR1 irrespective of redshift or model. 
As the original targeting for VUDS used an official COSMOS photometric catalogue (v1.7), the cross-matching of this catalogue with the photometric 
catalogue used for the fitting process is straightforward.

For ECDFS, in VUDS--DR1 we provide two different SED--fitting results based on two different photometric catalogues. The first
fitting was performed on the ground--based $U, B, V, R, I, z, J, H, K$ MUSYC photometry and \emph{Spitzer}/IRAC $[3.6], [4.5], [5.8], [8.0]$
photometry presented in \cite{Cardamone:10}, which was, in turn drawn from a variety of different surveys \citep[see Table 3 of ][]{Cardamone:10}.
Details of detection, PSF--homogenization, and correction of aperture magnitudes to close--to--total magnitudes is described in detail
in \citet{Cardamone:10}. The \cite{Cardamone:10} photometric source catalogue was used to select targets for VUDS and thus no additional
astrometry mapping was needed. The second set of fitting on ECDFS was performed on the U, F435W,
$F606W$, $F775W$, $F814W$, $F850LP$, $F098M$, $F105W$, $F125W$, $F160W$, $K_{s}$, [3.6], [4.5], [5.8], [8.0] mostly space-based photometry presented in \cite{Guo13}.
For further details on the properties and creation of this catalogue as well as how the CANDELS sources were matched and consistently photometered with the
ancillary data see \cite{Guo13}. VUDS--DR1 objects were mapped to the CANDELS catalogue through nearest--neighbour matching. We did not notice
any systematic offset of the derived physical parameters between the results of the SED fitting on the CANDELS photometry and the MUSYC photometry,
and the scatter of estimated physical parameters between the two sets of fits was consistent with Gaussian statistics. While both photometric
catalogues resulted in a large percentage of converging SED fits using Le Phare ($>97$\% for objects with measured magnitudes), we adopt for the presentation of VUDS--DR1
here the fitting based on the MUSYC photometry as it more closely matches the depth and filter set currently used in the COSMOS fitting.

For all sources in VUDS--DR1 with a measured spectroscopic redshift, these redshifts were used as a prior for the SED--fitting process. In the
fitting we employed seven different Bruzual \& Charlot (2003; hereafter BC03) models, five with star formation histories (SFHs) characterized by
an exponentially decaying tau model of the form $\psi(t)\propto \tau^{-1}e^{-t/\tau}$ and two delayed exponentially decaying tau models of the
form $\psi(t)\propto \tau^{-2}te^{-t/\tau}$. All templates are considered at 43 possible ages between 50 Myr and 13.5 Gyr with the
constraint that the age of the model fit to a given galaxy cannot be older than the age of the universe 
$t_{H}$ at the redshift of that galaxy. The delayed tau models are included as it has been suggested that high--redshift galaxies have SFHs which
may deviate considerably from the simple exponentially decaying tau models \citep[e.g.][]{Maraston10,Schaerer13}. Values of $\tau$ range from 0.1 to
30 Gyr in roughly logarithmically equal time steps and delay times are set to 1 and 3 Gyr. Each BC03 model employs a \cite{Chabrier:03} initial
mass function and two values of stellar--phase metallicity, 0.4$Z_{\odot}$ and $Z_{\odot}$. Stellar extinction is allowed to vary between
$E_{s}(B-V)=0$ and 0.5 in steps of 0.05, with the prescription also allowed to vary between the \cite{Calzetti:00} starburst law and a Small
Magellanic Cloud--like law \citep{Prevot:84,Arnouts:13}. Several prominent nebular emission lines are also added to the templates following the
methodology of \cite{Ilbert:09}. 

\subsection{UV luminosity, stellar masses and star formation rates}

Galaxies in the VUDS DR1 show a range of physical properties (e.g. stellar mass, SFR, dust extinction) comparable to that of the complete VUDS sample \citep{LeFevre:15}. This indicates that the VUDS DR1 forms a representative sample of the UV--selected star--forming galaxy population.

\subsection{Lyman--$\alpha$ EW}
The equivalent width (EW) of the Ly$\alpha$ line was measured manually using the \texttt{splot} tool in the \texttt{noao.onedspec} package in IRAF, similarly to \citet{Tresse:99}. We first put each galaxy spectrum in its rest--frame according to the spectroscopic redshift. 
Then, two continuum points bracketing the Ly$\alpha$ are manually marked and the rest--frame equivalent width is measured. 
The line is not fitted with a Gaussian, but the flux in the line is obtained integrating the area encompassed by the line and the continuum. 
This method allows the measurement of lines with asymmetric shapes (i.e., with deviations from Gaussian profiles), which is expected to be the case for most Ly$\alpha$ lines. 
The interactive method also allows us to control by eye the level of the continuum, taking into account defects that may be present around the line measured. 
It does not have the objectivity of automatic measurements, but, given the sometimes complex blend between Ly$\alpha$ emission and Ly$\alpha$ absorption, it does produce reliable and accurate measurements.

\begin{figure}
   \centering
   \includegraphics[width=\hsize]{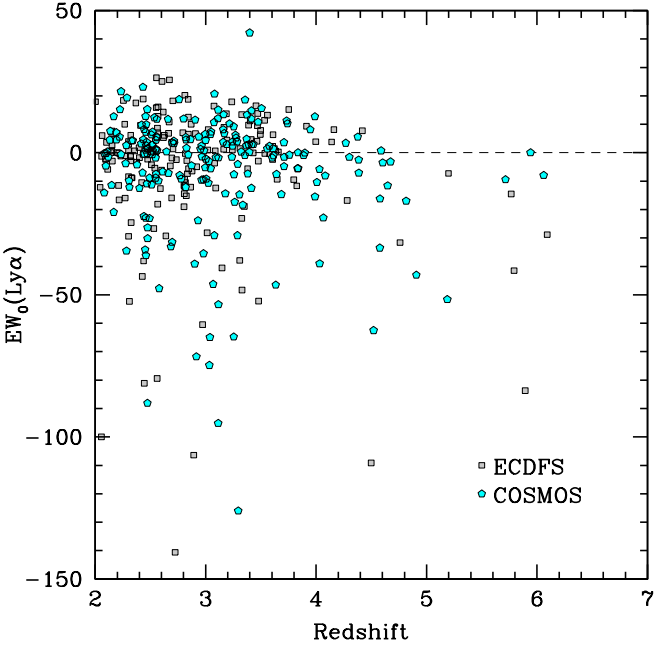}
      \caption{Lyman-$\alpha$ rest--frame equivalent widths as a function of spectroscopic redshift. Negative EW values indicate Ly$\alpha$ in emission. Galaxies in the VUDS--DR1 sample show predominantly Ly$\alpha$ in absorption up to z$\sim$3.5, and Ly$\alpha$ in emission at z$>$3.5 \citep[see ][for the full VUDS sample]{Cassata:15}.}  
         \label{lya_ew}
   \end{figure}

The distribution of rest--frame Lyman--$\alpha$ EW$_0$ as a function of redshift is presented in Figure \ref{lya_ew}. The VUDS--DR1 sample, like the full VUDS sample, is dominated by galaxies with Ly$\alpha$ in absorption for z$<$3.5 and Ly$\alpha$ in emission for z$>$3.5.

\section{Individual objects of interest}
\label{objects}

The VUDS sample in the CANDELS areas is rich of a number of interesting objects. We mentioned several over--dense regions in Sect. \ref{nz_dist}. Following \cite{Tasca:14} a number of galaxy pairs are identified from the secure spectroscopic redshift measurements of each component in a pair and will be used to calculate the pair fraction at these redshifts \citep{LeFevre:16}. 

To illustrate the content of the VUDS--DR1, we choose to present the 10 most massive galaxies verifying a redshift dependent stellar mass $log(M_{\star}) > -0.204\times(z - 5) + 10.15$. The evolution of this mass selection is based on the evolution of the characteristic mass as measured from the stellar mass function of \citet{Ilbert:13}, and our selection uses only the most reliable spectroscopic flags 3 and 4 (see Section \ref{sect_data} and Table \ref{flags_list} for a description of the spectroscopic flag system).

\begin{figure*}
   \centering
   \includegraphics[width=18cm]{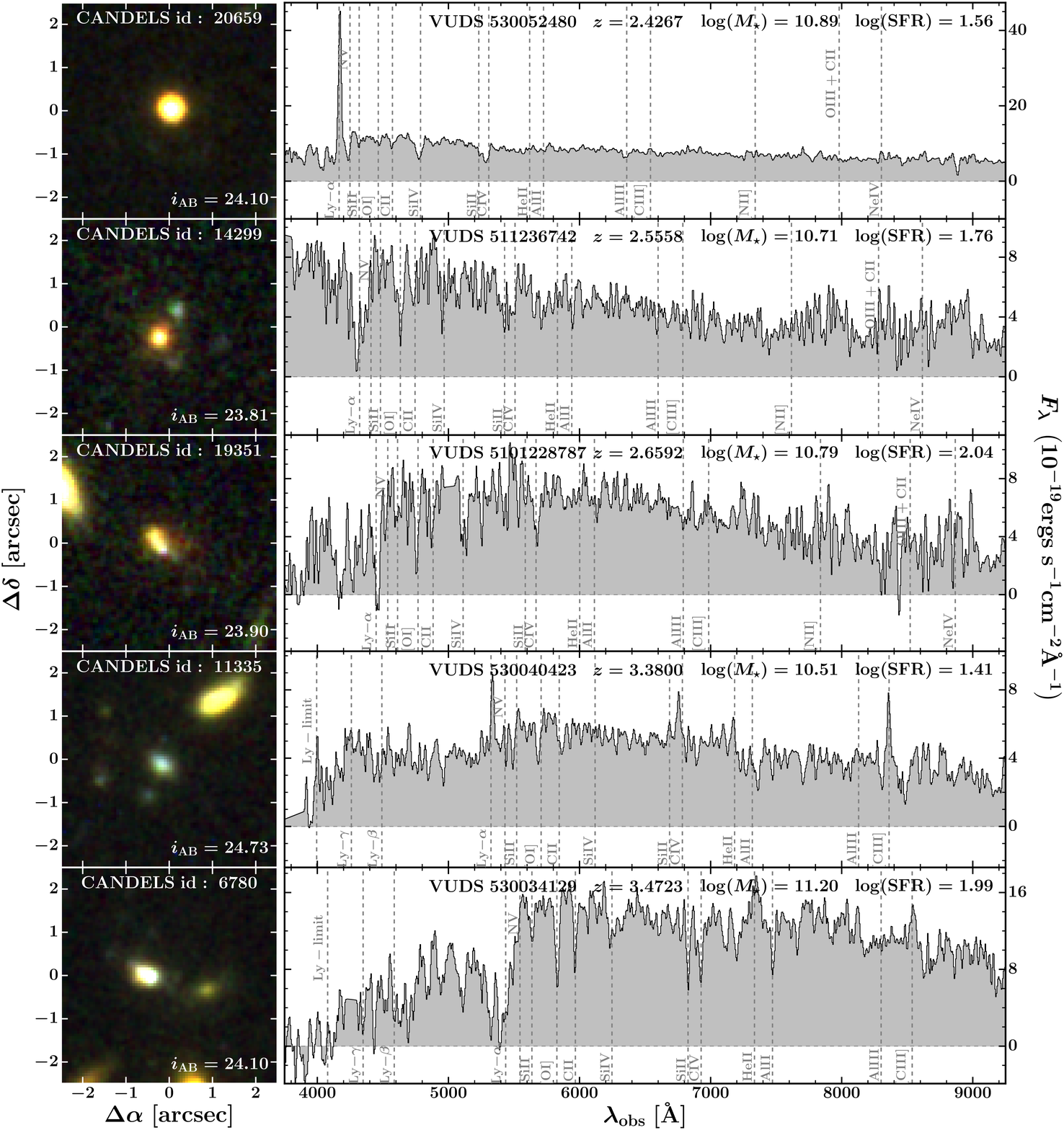}
      \caption{HST images, composite of F814W (blue), F125W (green), and F160W (red) from the CANDELS survey (left panels), and observed spectra from VUDS (right panels) for the ten most massive and stronger star--forming galaxies (see text) in the VUDS--DR1. The CANDELS ID number is indicated on each of the left panels, while the ID, the spectroscopic redshift from VUDS and the stellar mass M$_{\star}$ and SFR are indicated for each galaxy on the right panels.  
               }  
         \label{massive1}
   \end{figure*}

\begin{figure*}
   \centering
   \includegraphics[width=18cm]{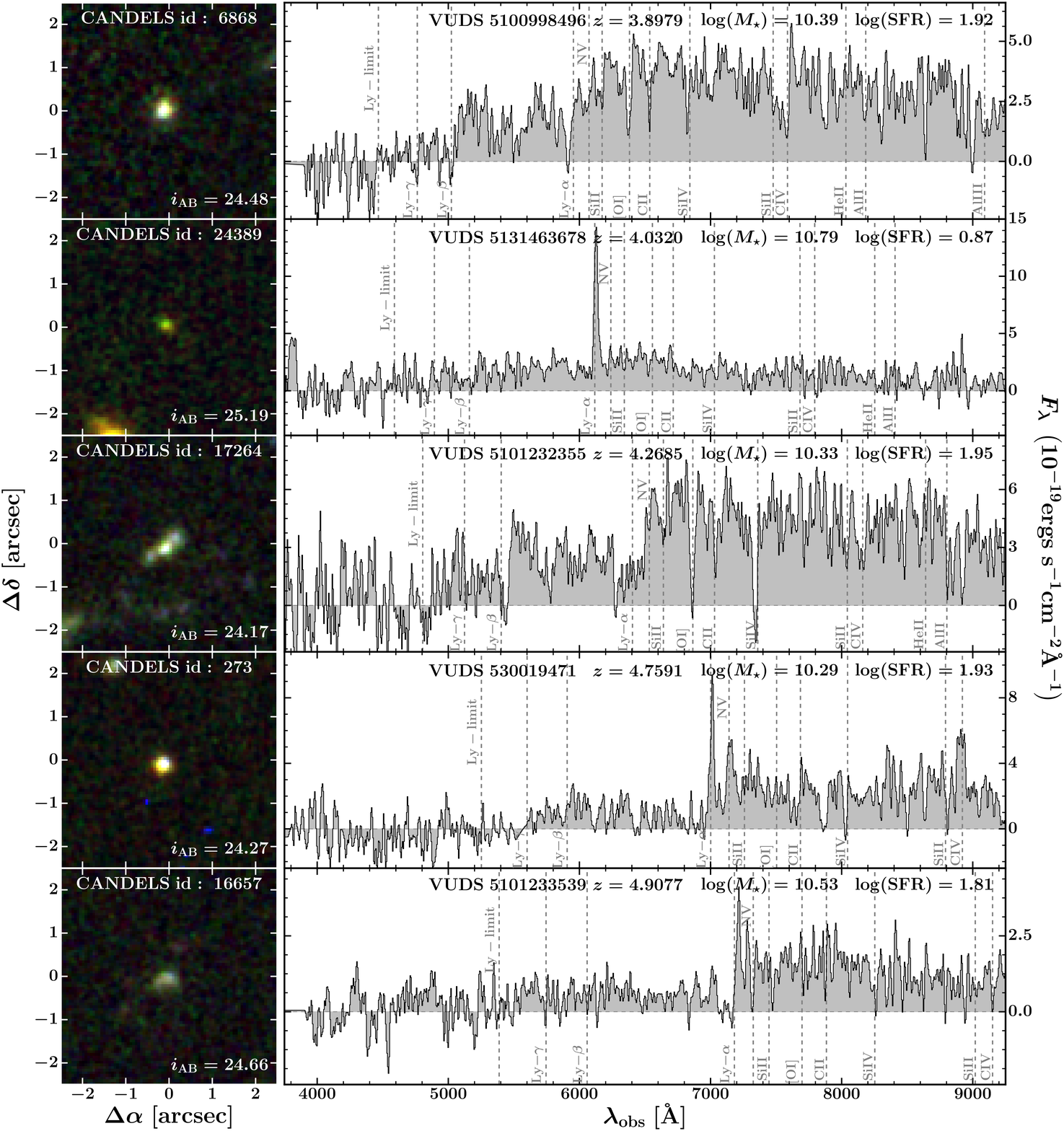}
      \caption{HST images, composite of F814W (blue), F125W (green), and F160W (red) from the CANDELS survey (left panels), and observed spectra from VUDS (right panels) for the ten most massive and stronger star--forming galaxies (see text) in the VUDS--DR1. The CANDELS ID number is indicated on each of the left panels, while the ID, the spectroscopic redshift from VUDS and the stellar mass M$_{\star}$ and SFR are indicated for each galaxy on the right panels.  
               }  
         \label{massive2}
   \end{figure*}

The properties of these ten most massive galaxies are listed in Table \ref{table_massive}. The VUDS spectra of these galaxies are presented together with the CANDELS images in Figure \ref{massive1} and \ref{massive2}. 
We find  massive galaxies at all redshifts from $z\sim2.5$ to $z\sim5$. The diversity of galaxy types at these masses is well illustrated in these examples with a large range of morphologies and spectral properties. Images show both very compact objects, and extended or multiple component objects, as well as pairs possibly indicative of mergers. The spectra of these objects include galaxies with only absorption lines as well as galaxies with strong Lyman--$\alpha$ emission. Their SFRs are generally large, ranging from $\sim7$ to more than 100 M$_{\sun}$/yr. 

\begin{table*}
\caption{The 10 most massive galaxies with log(M$_{\star}) > -0.204\times(z - 5) + 10.15$ in the VUDS--DR1}
\label{table_massive}    
\begin{tabular}{cccccccccc}       
\hline\hline                
VUDS--ID     & CANDELS--ID    &   $\alpha_{2000}$  & $\delta_{2000}$  & i$_{AB}$ & z$_{spec}$$^1$  & z$_{flag}$$^2$  & Age (Gyr)$^3$ & log(M$_{\star}$)$^4$  & log(SFR)$^5$ \\ \hline
5100998496 & 6868   & 150.20502  & 2.26216  & 24.48 & 3.8979 & 4 & 0.44  & 10.39 & 1.92   \\
5101228787 & 19351  & 150.15880  & 2.41091  & 23.89 & 2.6592 & 3 & 0.95  & 10.79 & 2.04  \\
5101232355 & 17264  & 150.18970  & 2.38569  & 24.17 & 4.2685 & 4 & 0.37  & 10.33 & 1.95  \\
5101233539 & 16657  & 150.18010  & 2.37834  & 24.66 & 4.9077 & 4 & 0.71  & 10.53 & 1.81  \\
511236742  & 14299  & 150.08801  & 2.35036  & 23.80 & 2.5558 & 4 & 1.42  & 10.71 & 1.76  \\
5131463678 & 24389  & 150.19586  & 2.48250  & 25.18 & 4.032  & 3 & 1.17  & 10.79 & 0.87  \\
530019471  & 273$^6$    &  53.12201  & -27.9387 & 24.26 & 4.7591 & 4 & 0.35  & 10.29 & 1.93  \\
530034129  & 6780   &  53.07440  & -27.8473 & 24.09 & 3.4723 & 4 & 1.28  & 11.20 & 1.99  \\
530040423  & 11335  &  53.06850  & -27.8071 & 24.73 & 3.38   & 3 & 1.31  & 10.51 & 1.41  \\
530052480  & 20659  &  53.18285  & -27.7349 & 24.10 & 2.4267 & 4 & 1.27  & 10.89 & 1.56  \\
\hline                                 
\end{tabular}
 \begin{list}{}{}
 \item[$^{\mathrm{1}}$] z$_{spec}$: spectroscopic redshift
 \item[$^{\mathrm{2}}$] z$_{flag}$: spectroscopic redshift reliability flag
 \item[$^{\mathrm{3}}$] Age, in Gyr: obtained from SED fitting at z$_{spec}$ 
 \item[$^{\mathrm{4}}$] log(M$_{\star}$): stellar mass (in log) obtained from SED fitting at z$_{spec}$ 
 \item[$^{\mathrm{5}}$] log(SFR): star formation rate (in log) obtained from SED fitting at z$_{spec}$ 
 \item[$^{\mathrm{6}}$] Identified as an AGN with an X-ray counterpart by \cite{Giallongo2015}
 \end{list}
\end{table*}

\section{Description of the VUDS--DR1 content}
\label{release}

\subsection{VUDS--DR1 data}
\label{sect_data}

The VUDS--DR1 includes the VUDS spectroscopic data in the CANDELS--COSMOS and CANDELS--ECDFS, as well as matched photometric data and derived quantities, as described below. 
 
The primary catalogue information concerns the VIMOS spectroscopic data. The fundamental advance provided by this release is the spectroscopic redshift z$_{spec}$. These measurements result from the data reduction described in Sect.\ref{obs} \citep[see][for more details]{LeFevre:05,LeFevre:13a,LeFevre:15}. From repeated observations of VUDS galaxies, as well as duplicate observations using the same configuration as in the VVDS \citep{LeFevre:13a}, the velocity accuracy of redshift measurements is about $\sigma_V=200$ km/s. A flagging scheme is associated to each redshift measurements. 
Originated with the Canada--France Redshift Survey \citep[CFRS,][]{LeFevre95}, this flag system indicates the \emph{reliability} of the redshift being accurate based on repeatability of the measurements on re-observed targets and does not speak to metrics of ``quality" of the spectrum from which the redshift is measured, a term which may manifest itself in a variety of definitions across different surveys \citep[see, e.g., discussion in ][]{LeFevre:13a}.
Besides VUDS, this flag system, or slight variants of it, have been used by the VVDS, zCOSMOS and VIPERS surveys \citep{LeFevre:13a,Lilly:07,Guzzo:14}. The VUDS flag system inspired from this is described in \cite{LeFevre:15} and a complete list with the meaning of the spectroscopic flags in the VUDS--DR1 is given in Table \ref{flags_list}. 

Associated to the spectroscopic redshift and reliability flag, the 2D and 1D object spectra and corresponding noise are made available (in FITS format). 
The 2D spectra are the result of the data processing described in Sect.\ref{obs}, after sky subtraction, stacking of the individual exposures, wavelength and flux calibration.  
Given a 2D spectrum,we can compute sky subtraction residuals as
\begin{equation}
\label{residual}
    \langle{\rm skyResidual}\rangle_{i}= \frac{\sum_{j \in {\rm NRegion}} (\langle C\rangle-C_j)^2}{{\rm NRegion}-1} ,
\end{equation}
where NRegion is the number of pixels in the sky region on one side of the object,
$\langle C\rangle$ is the mean of the counts in the sky region and $C_j$ are the counts in pixel $j$,
at wavelength $i$.
This computation is done on both sides of the object (resL and resR),
taking into account slit borders and possible second objects in the slit.
From such residuals, a mean noise spectrum is computed,
the noise in the $i$-th pixel being given by
\begin{equation}
\label{noiseSpectrum}
   {\rm noise}_{i}^2=\frac{{\rm resL}_i+{\rm resR}_i}{2}+S_i=\langle{\rm skyResidual}\rangle_i + S_i ,
\end{equation}
where $S_i$ are the source counts at wavelength $i$.

The 1D spectra are produced from joining the 1D spectra extracted from the LRBLUE and LRRED 2D spectrograms, and are calibrated in wavelength and flux.
To join the VIMOS blue and red grism spectra which have a slightly different spectra dispersion (5.3 and 7.5 \AA/pixel), the red spectrum is re-sampled to the same dispersion as the blue one. The overlapping region between 5600 and 6700 \AA \space is used to compute the flux normalization factor between the two spectra. The S/N ratio of the final joined spectrum is the the S/N weighted mean of the two spectra in the overlapping region.
The 1D spectra are corrected for atmospheric effects (refraction and extinction) comparing with broad band photometry from the u--band to the i--band. The pseudo--spectrum produced by atmospheric refraction at the elevation of the observations, made available in the VIMOS spectra headers, is computed in combination with an elliptical geometrical model of each galaxy image to estimate the fraction of the light lost through the one-arcsecond wide slit \citep{Thomas:14}. This correction concerns mostly the bluer spectral data with $\lambda \leq 4500$\AA. After this process we find that the agreement between the spectra flux and the broad band photometry is accurate to better than 10\% r.m.s for $\lambda \leq 4500$\AA \space and 5\% r.m.s. for $\lambda > 4500$\AA.

\subsection{Cross--match with the HST--CANDELS catalogue}

Each of the VUDS catalogue entries were cross--matched to the CANDELS source list in each of the COSMOS and ECDFS fields as kindly provided by the CANDELS collaboration. For each VUDS target the coordinates based on the original VIMOS targeting scheme were cross--matched to the CANDELS catalogue using a matching radius of 0.5$\arcsec$. As a result of this process, all of the VUDS entries are associated to a CANDELS identification number. This facilitates data searches across different catalogues. When a source is resolved in the HST images with multiple components identified in the CANDELS catalogue, only the closest match to the VUDS catalogue is chosen. Therefore while a single VUDS source can have several counterparts in CANDELS, the reverse should not happen, and we estimate that this concerns 1.4\% of objects for a 0.5 arcsecond search radius (this would rise to 12.3\% for a 1 arcsecond search radius). Such a resolution effect may have important implications e.g. in counting faint galaxies, and we mention it here to alert the reader.

All spectroscopic data and parameters derived from SED fitting are available in the VUDS--DR1. We list the main parameters of interest in Table \ref{parameters}.

\begin{table*}
\caption{VUDS redshift reliability flags summary}
\label{flags_list}    
\begin{tabular}{ccl}       
\hline\hline                
z$_{flag}$  & Reliability                                   & Comments \\ \hline
4           & 100\%                                         & High S/N with many absorption and/or emission lines; strong cross--correlation signal with excellent \\
            &                                               & continuum match to templates        \\
3           & 95-100\%                                      & Moderate to high S/N  with several absorption and/or emission lines; strong cross--correlation signal with good to \\ 
            &                                               & excellent continuum match to templates    \\
2           & 70-80\%                                       & Moderate S/N with matching absorption and/or emission lines; \\
            &                                               & good cross--correlation signal      \\
1           & 40-50\%                                       & Low S/N lines or continuum; weak to moderate cross--correlation signal  \\
9           & 80 \%                                         & Single emission line spectrum; the line has moderate to high S/N. \\
            &                                               & Redshift is assigned based on the exclusion of alternative solutions  \\
            &                                               & in the absence of other expected  key features \\
1X          & --                                            & Like flag X but spectral features indicate the presence of an AGN \\
2X          & --                                            & Like flag X but for a secondary target falling in the slit of a primary \\
3X          & --                                            & Like flag X but the spectrum shows two line systems at different redshifts, the highest of the two redshifts is given as z$_{spec}$ \\
4X          & --                                            & Like flag X but the two spectra are close in velocity indicative of a physical pair \\
\hline                                 
\end{tabular}
\end{table*}

\begin{table*}
\caption{VUDS--DR1 database parameters}
\label{parameters}    
\begin{tabular}{ccccccc}       
\hline\hline                
Parameter   & Description                                   & Comments \\ \hline
VUDS--ID     & VIMOS Ultra--Deep Survey Identification number &        \\
CANDELS--ID  & CANDELS Identification number                 &        \\
z$_{spec}$  & Spectroscopic redshift measured with VIMOS    & Typical accuracy: $\Delta$z=0.0007, $\lesssim$200 km/s \\
z$_{flag}$  & Spectroscopic redshift reliability            &        \\
EW(Ly$\alpha$) & Rest--frame Ly$\alpha$--1215\AA ~equivalent width       &        \\
M$_{\star}$  &  Stellar mass                                 & From SED fitting at z$_{spec}$      \\
SFR         &  Star formation rate                          & From SED fitting at z$_{spec}$      \\ 
Age         &  Mean age of the stellar population corresponding to the best fit &  From SED fitting at z$_{spec}$      \\
E(B-V)      &  Galactic extinction                          & From SED fitting at z$_{spec}$      \\ 
L$_{NUV}$   &  Absolute luminosity at 1500\AA \space rest--frame    & From SED fitting at z$_{spec}$      \\
M$_{band}$  &  Absolute magnitudes in {\it band}            & From SED fitting at z$_{spec}$      \\ 
\hline                                 
\end{tabular}
\end{table*}


\section{Summary}
\label{summary}

The VIMOS Ultra Deep Survey first Data Release (VUDS--DR1) gives full access to  low--resolution spectroscopic data of 698 sources in the CANDELS--COSMOS and CANDELS--ECDFS fields, covering a total area  of 276.9 arcmin$^2$. The key new information provided by this release is the spectroscopic redshifts $z_{spec}$, and associated reliability flag, as well as the full 1D flux, the wavelength calibrated VIMOS spectra, and the Liman--$\alpha$ rest--frame equivalent widths. 
We release also SED--derived quantities at the spectroscopic redshift value, including stellar mass M$_{\star}$ and star formation rate. 
Out of the 677 galaxies identified about 500 are at a redshift $z_{spec}>2$ and up to $z_{spec}\sim6$, with $\sim50$ with $z_{spec}>4$. The VUDS--DR1 therefore approximately doubles the number of high redshift galaxies with spectroscopic redshifts measured at $z>3$ in the COSMOS and ECDFS fields.

All the 698 objects in the CANDELS--COSMOS and CANDELS--ECDFS are publicly released from the CESAM data center at \texttt{http://cesam.lam.fr/vuds/DR1}. 

Future VUDS data releases will follow this VUDS--DR1 to give complete access to the spectra and associated measurements of $\sim$8\,000 objects in the full $\sim$1 square degree of the VUDS survey.


\begin{acknowledgements}
This work is supported by funding from the European Research Council Advanced Grant ERC--2010--AdG--268107--EARLY and by INAF Grants PRIN 2010, PRIN 2012 and PICS 2013. 
AC, OC, MT and VS acknowledge the grant MIUR PRIN 2010--2011.  
This work is based on data products made available at the CESAM data center, Laboratoire d'Astrophysique de Marseille, France. 
\end{acknowledgements}


\bibliographystyle{aa} 
\bibliography{first_release} 


\end{document}